\documentclass[12pt,a4paper]{article}
\usepackage{epsfig}
\usepackage{amstex}

\catcode`\@=11
\long\def\@makefntext#1{
\protect\noindent \hbox to 3.2pt {\hskip-.9pt  
$^{{\ninerm\@thefnmark}}$\hfil}#1\hfill}		

\def\@makefnmark{\hbox to 0pt{$^{\@thefnmark}$\hss}}  
	
\def\ps@myheadings{\let\@mkboth\@gobbletwo
\def\@oddhead{\hbox{}
\rightmark\hfil\ninerm\thepage}   
\def\@oddfoot{}\def\@evenhead{\ninerm\thepage\hfil
\leftmark\hbox{}}\def\@evenfoot{}
\def\sectionmark##1{}\def\subsectionmark##1{}}

\renewcommand{\thefootnote}{\fnsymbol{footnote}}

\newcounter{sectionc}\newcounter{subsectionc}\newcounter{subsubsectionc}
\renewcommand{\section}[1] {\vspace*{0.6cm}\addtocounter{sectionc}{1} 
\setcounter{subsectionc}{0}\setcounter{subsubsectionc}{0}\noindent 
	{\normalsize\bf\thesectionc. #1}\par\vspace*{0.4cm}}
\renewcommand{\subsection}[1] {\vspace*{0.6cm}\addtocounter{subsectionc}{1} 
	\setcounter{subsubsectionc}{0}\noindent 
	{\normalsize\it\thesectionc.\thesubsectionc. #1}\par\vspace*{0.4cm}}
\renewcommand{\subsubsection}[1]
{\vspace*{0.6cm}\addtocounter{subsubsectionc}{1}
	\noindent {\normalsize\rm\thesectionc.\thesubsectionc.\thesubsubsectionc. 
	#1}\par\vspace*{0.4cm}}

\newcounter{appendixc}
\newcounter{subappendixc}[appendixc]
\newcounter{subsubappendixc}[subappendixc]

\renewcommand{\appendix}[1] {\vspace*{0.6cm}
        \refstepcounter{appendixc}
        \setcounter{figure}{0}
        \setcounter{table}{0}
        \setcounter{equation}{0}
        \renewcommand{\thefigure}{\Alph{appendixc}.\arabic{figure}}
        \renewcommand{\thetable}{\Alph{appendixc}.\arabic{table}}
        \renewcommand{\theappendixc}{\Alph{appendixc}}
        \renewcommand{\theequation}{\Alph{appendixc}.\arabic{equation}}
        \noindent{\bf Appendix \theappendixc #1}\par\vspace*{0.4cm}}

\def\abstracts#1{{
	\centering{\begin{minipage}{12.2truecm}\footnotesize\baselineskip=12pt\noindent
	\centerline{\footnotesize ABSTRACT}\vspace*{0.3cm}
	\parindent=0pt #1
	\end{minipage}}\par}} 

\renewenvironment{thebibliography}[1]
	{\begin{list}{\arabic{enumi}.}
	{\usecounter{enumi}\setlength{\parsep}{0pt}
\setlength{\leftmargin 1.25cm}{\rightmargin 0pt}
	 \setlength{\itemsep}{0pt} \settowidth
	{\labelwidth}{#1.}\sloppy}}{\end{list}}

\newcounter{itemlistc}
\newcounter{romanlistc}
\newcounter{alphlistc}
\newcounter{arabiclistc}

\newcommand{\fcaption}[1]{
        \refstepcounter{figure}
        \setbox\@tempboxa = \hbox{\footnotesize Fig.~\thefigure. #1}
        \ifdim \wd\@tempboxa > 6in
           {\begin{center}
        \parbox{6in}{\footnotesize\baselineskip=12pt Fig.~\thefigure. #1}
            \end{center}}
        \else
             {\begin{center}
             {\footnotesize Fig.~\thefigure. #1}
              \end{center}}
        \fi}

\newcommand{\tcaption}[1]{
        \refstepcounter{table}
        \setbox\@tempboxa = \hbox{\footnotesize Table~\thetable. #1}
        \ifdim \wd\@tempboxa > 6in
           {\begin{center}
        \parbox{6in}{\footnotesize\baselineskip=12pt Table~\thetable. #1}
            \end{center}}
        \else
             {\begin{center}
             {\footnotesize Table~\thetable. #1}
              \end{center}}
        \fi}

\def\@citex[#1]#2{\if@filesw\immediate\write\@auxout
	{\string\citation{#2}}\fi
\def\@citea{}\@cite{\@for\@citeb:=#2\do
	{\@citea\def\@citea{,}\@ifundefined
	{b@\@citeb}{{\bf ?}\@warning
	{Citation `\@citeb' on page \thepage \space undefined}}
	{\csname b@\@citeb\endcsname}}}{#1}}

\newif\if@cghi
\def\cite{\@cghitrue\@ifnextchar [{\@tempswatrue
	\@citex}{\@tempswafalse\@citex[]}}
\def\citelow{\@cghifalse\@ifnextchar [{\@tempswatrue
	\@citex}{\@tempswafalse\@citex[]}}
\def\@cite#1#2{{$\null^{#1}$\if@tempswa\typeout
	{IJCGA warning: optional citation argument 
	ignored: `#2'} \fi}}

 1
 1
 1

\font\ninerm=cmr9

\newcommand{\epem}{\mbox{$e^+e^-$}}
\newcommand{\Pj}{\mbox{$P_{j}$}}
\newcommand{\eref}[1]{(\ref{#1})}
\def\nostrocostrutto#1\over#2{\mathrel{\mathop{\kern 0pt \rlap 
  {\raise.2ex\hbox{$#1$}}}
  \lower.9ex\hbox{\kern-.190em $#2$}}}
\def\gsim{\nostrocostrutto > \over \sim}   

\begin{document}
\thispagestyle{empty}

\centerline{\normalsize\bf RECENT TESTS OF PARTON HADRON DUALITY}
\centerline{\normalsize\bf IN MULTIPARTICLE FINAL STATES\footnote{
MPI-PhT/97-52, August 1997, To be publ. in Proc. of the Ringberg workshop
\lq\lq New Trends in HERA Physics", Tegernsee, Germany, 25-30 May 1997.}}

\vspace*{0.6cm}
\centerline{\footnotesize WOLFGANG OCHS}
\baselineskip=13pt
\centerline{\footnotesize\it Max-Planck-Institut f\"ur Physik
(Werner-Heisenberg-Institut)}
\baselineskip=12pt
\centerline{\footnotesize\it F\"ohringer Ring 6, D-80805 Munich, Germany}
\centerline{\footnotesize E-mail: wwo@mppmu.mpg.de}
\vspace*{0.9cm}
\abstracts{
A large variety of jet properties can be described
perturbatively by evolving the parton cascade down to small scales of order
of a few 100 MeV. We discuss two recent applications of this approach:
1. the soft limit of the particle spectrum which is nearly energy
independent as expected for the soft gluon bremsstrahlung; further tests of
this interpretation in $e^+e^-$ and DIS are presented; 2. the steep decrease
of the distribution of rapidity gaps in  $e^+e^-$ annihilation can be
explained by the Sudakov formfactor for quark jets. }
\normalsize\baselineskip=15pt
\setcounter{footnote}{0}
\renewcommand{\thefootnote}{\alph{footnote}}

\section{Introduction}
An important goal in the study of high energy hard collision processes
is the detailed understanding of the 
 hadronic final state and its characteristic jet structure. 
There are two mayor fields of interest, one concerning the predictions in
QCD perturbation theory, mainly on the jet final states,
the second concerning the colour
confinement phenomena, how the partonic cascade evolves
into the final state of hadrons.

In the popular models for particle production in 
hard collision processes  it is assumed that first 
the primarely produced partons evolve by bremsstrahlung processes
according to perturbation theory into jets of partons until a characteristic
resolution scale of $Q_0\sim 1$ GeV is reached.
Thereafter, non-perturbative
processes take over and the final hadronic particles, often through
intermediate resonances, are produced, for example, by a string
mechanism\cite{JETSET}
or through  cluster formation.\cite{HERWIG} 
These models have been proven to be very successful
in the description of the experimental data over the years. On the other
hand, they are only accessible through a Monte Carlo code and involve 
quite a number of a priori unknown parameters in the description
of the hadronization phase.

We consider here another approach, 
based on the concept of ``Local Parton Hadron
Duality'' (LPHD).\cite{adkt}
The parton cascade is evolved further
down to a scale of about $Q_0\sim 250$ MeV 
where $Q_0$ is now the cut-off in the transverse momentum.
Then one compares directly an observed quantity with the 
corresponding parton level calculation 
without any hadronization correction. 

An immediate benefit of this approach are the analytical
results in many cases which provide a deeper understanding of
how the theory is
actually tested by a particular measurement.
For example, 
the  scaling properties and the systematics of their violations 
can only be formulated in a meaningful way using analytical formulae.
Some results can be expressed in compact form by
simple QCD numbers.

Theoretical calculations can be carried out in the simplest case in the
Double Log Approximation (DLA),\cite{dfk1,bcm}
which provides the high energy limit, or in the Modified Leading
Log Approximation (MLLA)\cite{MLLA}
which includes finite energy corrections; they are usually essential
to obtain quantitative agreement with experiment at present
energies. 
The original success was in the description of the inclusive hadron
distributions,\cite{adkt} 
thereafter it
has been applied to a large variety of different observables.\cite{dkmt}   
In recent years more applications have been worked out and the
experimental results from LEP, HERA and TEVATRON gave further
support to this approach.\cite{ko}

The duality picture is clearly rather bold and it is not clear a priori
for which observables and in which kinematic ranges it applies.
This has to be clarified ultimately by experiment.
It is obvious that this model cannot compete with
the standard hadronization models in the description of 
the various details of the
final state like the production of different 
particle species or resonances. It has so far been applied successfully
for suitably averaged inclusive quantities.

The main virtue of this approach is its intrinsic simplicity which involves
only two
essential parameters, namely the QCD scale $\Lambda$ and the 
non-perturbative mass scale $Q_0$.
In some cases the absolute normalization is taken as additional parameter;
on the other hand, in a recent calculation of multiplicities
beyond MLLA it was suggested 
that one parton corresponds to one  
hadron in this duality picture.\cite{lo2}

The hope is that with increased phenomenological knowledge about successes
and limitations of this approach our theoretical 
understanding of the confinement mechanism will improve.
In this presentation we  dicuss two recent applications,
one concerning the soft limit of the energy spectrum\cite{klo1}
and the other one the rate for events with 
large rapidity gaps,\cite{os} i.e. events in a quasi exclusive limit.

\section{Main Ingredients of the Analytical Calculations}
In this section we give a short overview of the analytical treatment of
multiparticle processes. This involves
the matrix elements for the primary hard subprocess and the 
evolution of the primary partons into  jets by bremsstrahlung cascades.
The  gluon bremsstrahlung at small angles $\vartheta$ with energy $E$ 
 off the primary parton of type $A$ ($A = q, g$)
with jet momentum $\Pj$ is given by 
\begin{equation}
  dn_A = \frac{C_A}{N_C} \gamma_0^2(p_\perp) \frac{d \vartheta}{\vartheta} \frac{dE}{E},
 \qquad
  \gamma_0^2(p_\perp) = \frac{2N_C \alpha_s(p_\perp)}{\pi}
                  = \frac{\beta^2}{\ln(p_\perp/\Lambda)}, 
  \qquad p_\perp \ge Q_0
\label{gluon}
\end{equation}
where $p_\perp \approx E \vartheta$, $\beta^2 = 4 N_C/b$,
$b \equiv (11 N_C - 2n_{f})$/3 with $N_C$, $n_f$
the numbers of colours and flavours, also $C_g$ = $N_C$, $C_q$ = 4/3.
Inside the cascade the soft gluons are coherently produced from all
harder partons. For azimuthally averaged quantities the consequences
of the coherence effect can be taken into account by the angular 
ordering prescription\cite{AO} which requires
 the angles of subsequent gluon emissions to be in decreasing order.

The multiparticle properties of the jet can be discussed
conveniently by using the generating functional
$Z_A(\Pj, \Theta;{u(p)})$. Here $\Pj$ and $\Theta$ denote the initial parton
momentum and 
opening angle of the jet, and $u(p)$ is a profile function for
particle momentum $p$. The functional is constructed from all
the exclusive final states. Then the inclusive densities
can be obtained by functional differentiation with respect to 
 the profile function
$u(p)$, for example the one particle density by
$ \rho^{(1)} (p)= \delta Z\{u\}/\delta u(p)
\mid_{u=1}$.
The properties of these densities can be derived from the
evolution equation for $Z$ which relates the functional
at scales $\Pj, \Theta$ to the one at lower scales according to the
``decay'' $A \to BC$. In MLLA accuracy this evolution equation
is given by\cite{dkmt}
\begin{multline}
\frac{d}{d \: \ln \: \Theta} \: Z_A 
 (\Pj, \Theta) = 
\frac{1}{2}
\; \sum_{B,C} \; \int_0^1 \; dz \\
\times 
\frac{\alpha_s (p_\perp^2)}{2 \pi} \: \Phi_A^{BC} (z)
[Z_B (z\Pj, \Theta) \: Z_C ((1 - z)\Pj, \Theta) \: - \: Z_A
(\Pj,\Theta) ]  \label{evz}
\end{multline}
where $\Phi_A^{BC}(z)$ denotes the DGLAP splitting functions.
The initial condition for the parton evolution is given by
\begin{equation}
Z_A (\Pj, \Theta; \{ u \})|_{P_j \Theta  = Q_0} \; = \; u_A
(p = \Pj),
\label{ic}
\end{equation}
i.e. at threshold there is only the primary parton.

From (\ref{evz}) one can obtain the evolution equations for
particle densities by appropriate functional differentiation. 
Therefore
this equation is the basic tool for deriving the multiparticle
properties of a jet analytically. A more detailed discussion on the
approximations can be found elsewhere.\cite{dkmt,ko,do,woz}

\section{The Soft Limit of the Particle Spectrum}
\subsection{Theoretical Predictions Confronted with Experiment}
An essential ingredient of the LPHD approach is the evolution of the parton
cascade towards small scales. 
The normalization of all quantities
is given by the initial condition (\ref{ic}).
Therefore, one has to assume that the perturbative evolution can be applied
at small  $cms$ energies as well.
Alternatively, one could start the evolution from a larger $cms$ energy
whenever the perturbative QCD is considered trustworthy, but then one had to
introduce unknown parameters for initialization instead of (\ref{ic})
and a lot of the predictive power of this approach would disappear.
Indeed, applying the MLLA calculations\cite{DKTInt} 
the moments of the energy spectra are well described  down to such 
low energies as $\sqrt s=3$ GeV using the initial condition
(\ref{ic}).\cite{lo}

Of special interest is the behaviour of the soft end of the particle energy
spectrum. 
The coherence of the soft gluon emission from all harder partons
forbids the multiplication of the soft particles\cite{adkt} 
and the spectrum
in the variable $\xi=\ln(\Pj/E)$ obtains a \lq\lq hump-backed''
shape.\cite{dfk1,bcmm}

This problem has been studied recently in more detail.\cite{klo1}
The parton density in the variable $\xi=\ln(\Pj/E)$ 
in the DLA is given in lowest order of $\alpha_s$ by 
\begin{equation}
\frac{dn_A(E,\Pj)}{d\ln E}=
\frac{C_A}{N_C}\beta^2\ln\frac{\ln(E/\Lambda)}{\ln(Q_0/\Lambda)}
\label{dxi}
\end{equation}
This single bremsstrahlung contribution is independent of the jet energy
$\Pj$; the higher order contributions do depend on $\Pj$ but they 
are negligable in the soft limit $E\to
Q_0$. Also the MLLA correction leaves 
this limit unaltered:
the energy conservation effects and large $z$ corrections from the splitting
functions which make up the differences between DLA and MLLA
can be neglected. 
If the perturbation theory and LPHD are valid towards
such low particle energies $E$ one expects then also that
the hadron spectrum becomes independent of the $cms$ energy
in the soft limit. To exhibit the energy dependence it is convenient
to consider 
 the invariant density $Edn/d^3p$
which remains finite for small momenta.\footnote{Another 
alternative is the
non-invariant spectrum  $dn/d^3p$\cite{klo2}
which has the advantage of being independent of the particle mass.}
~$\,$Then we expect an energy independent 
density $I_0$
of hadrons in the limit where the particle momentum
$p$, or alternatively 
rapidity $y$ and transverse momentum $k_\perp$, become small:
\begin{equation}        
I_0 = \lim_{y \to 0, p_T \to 0} E \frac{dn}{d^3p} \quad = \quad
\frac{1}{2} \lim_{p \to 0} E \frac{dn}{d^3p}.
\label{izero}
\end{equation}        
The factor $\frac{1}{2}$ takes into account that both
hemispheres are included in the limit $p \to 0$. 

This prediction
is a direct consequence of the coherence of the soft gluon
emission: the emission rate for the gluon of large wavelength
does not depend on the details of the jet evolution at smaller distances;
it is essentially determined by the colour charge of the hard initial
partons.

\begin{figure}[t]
          \begin{center}
\mbox{\epsfig{file=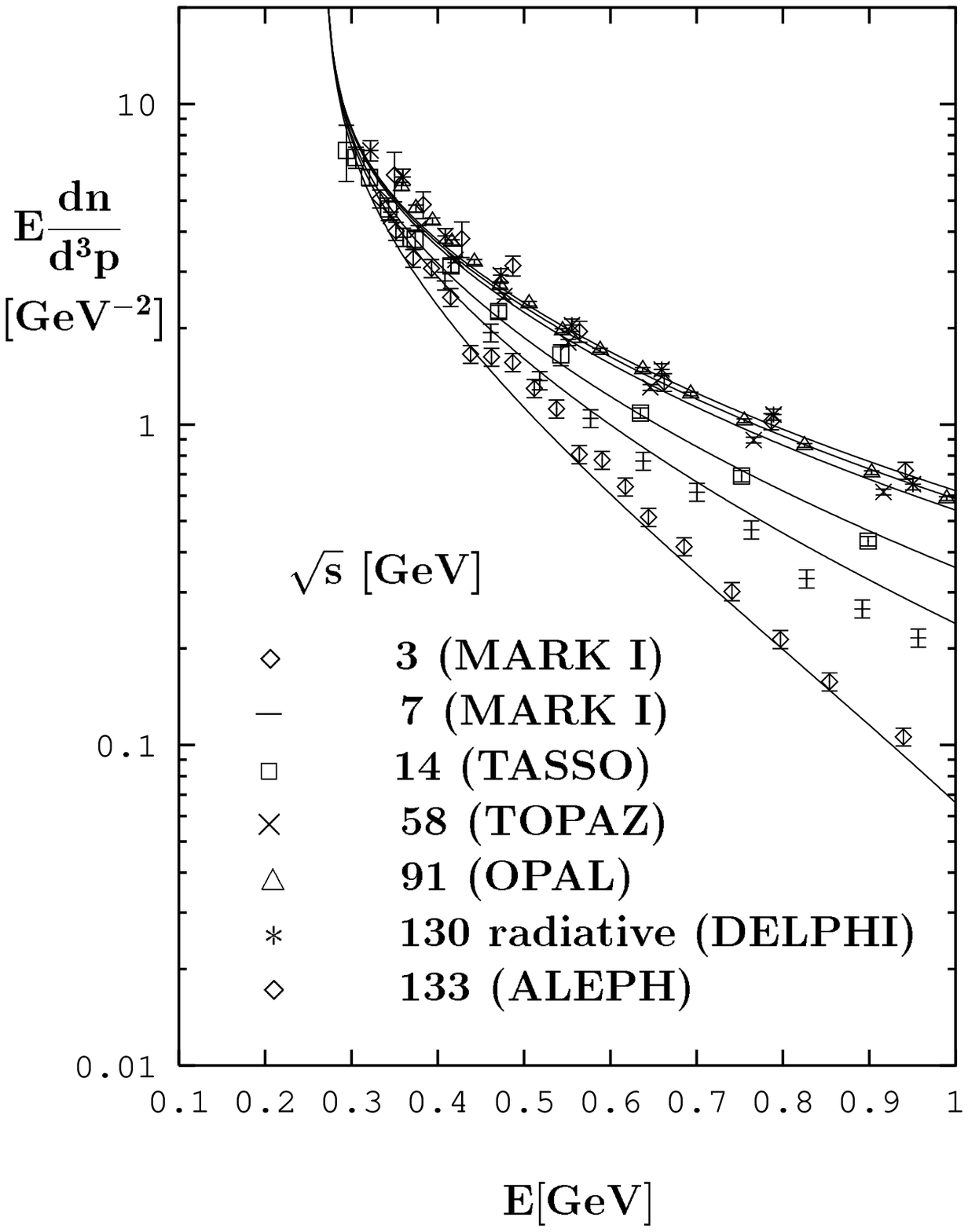,width=6.6cm,bbllx=4.5cm,bblly=9.5cm,bburx=16.5cm,bbury=26.cm}}
       \end{center}
\fcaption{Invariant density $E dn/d^3p$ of charged particles
in $e^+e^-$ annihilation
as a function of the particle energy
$E=\protect\sqrt{p^2+Q_0^2}$ at $Q_0$ = 270 MeV.
Experimental data  at various $cms$ energies
from  SLAC, TASSO and TOPAZ Collaborations, LEP-1 and LEP-1.5
are compared to MLLA predictions
(taken from Ref. 10).}
\label{chargedall}
\end{figure}

In Fig.~1 we show the experimental results on the invariant density
of charged particles
for $cms$ energies from 3 to 130 GeV in $e^+e^-$ annihilation.
An approximate energy independence of the low momentum particle
density (within about 20\%) is
indeed observed; the same is true for identified particles $\pi$,
$K$ and $p$.\cite{klo1}
The curves in Fig.~1 represent the MLLA results. 
At very low energies where $E\approx Q_0$ the QCD perturbative results have
to be supplemented by a kinematical relation between parton and hadron
spectra taking into account their
different masses
(for more details, see Ref. 10). 
The theoretical curves show the approach to the scaling limit and
describe well the different slopes at larger particle energies. An important
role here is played by the running $\alpha_s$ which provides the
strong rise towards small energies for $E<1$ GeV, for fixed $\alpha_s$ this
rise would be much weaker.\cite{lo,klo1}

Recently inclusive spectra from HERA in the current fragmentation region
in the Breit frame became available.\cite{H1} The data are found to 
be rather similar in shape to the observation in \epem annihilation
at comparable scales.
Especially, the approximate energy independence of the soft part of the
spectrum is verified in the region $12<Q^2<100$ GeV$^2$. This supports the
universality of the low energy phenomena. 
 
\subsection{Further Tests of the Perturbative Picture}
One may suspect that the universal, approximately 
energy independent soft limit of the invariant spectrum could be some general
hadronization phenomenon  not related to the coherence properties of
perturbative QCD. It is therefore very important to further investigate the
perturbative origin of the effect.
Whereas there is no parameter free prediction of the absolute
size of the density $I_0$, the perturbative approach predicts its dependence
on the colour charge of the primary parton as in (\ref{dxi}). Especially,
we expect for the ratio of soft particle
densities  in a $gg$ system to the one in a
$q\bar q$ system 
\begin{equation}
\frac{I_0^g}{I_0^q} =\frac{N_C}{C_F}=\frac{9}{4}.
\label{rgq}
\end{equation}
Such a ratio is expected asymptotically for the total multiplicities of
particles,\cite{bg} but at present energies there are large corrections.
On the other hand, such a large ratio may nevertheless appear
in case of the soft
particles where the DLA asymptotic behaviour is valid to a good
approximation.
The preparation of $gg$ final states is not easy; but there are two lines of
approach which can test the hypothesis (\ref{rgq}),\cite{klo1,klo2}
the first at a more qualitative, the second at a quantitative level.\\

\noindent{\it Study of the central rapidity plateau at small $p_\perp$}\\
Let's first consider \epem annihilation. The rate $I_0$ in (\ref{izero}) 
is defined in the
$cms$. If we perform a Lorentz transformation in direction of one primary
parton, we arrive at a frame with unequal momenta of $q$ and $\bar q$,
say at rapidity $y$ from the $cms$. 
As the soft radiation does not depend on the primary energy, 
the density $I_0(y)$ in the frame
at $y$ has not changed, so $I_0(y)$ should be constant away from the
kinematical boundaries. Instead of the density at $p_\perp=0$ we can also
consider the density in an interval at small $p_\perp$.  
This picture is a bit 
idealistic, in that the primary parton direction is not known and should be
replaced by a suitable jet-axis, then $I_0(y)$ will not be exactly constant.   

Next we consider collisions with incoming hadrons or/and photons.
The interaction may proceed through quark or gluon exchange where we have to
assume some semihard momentum transfer of, say, at least 
1~GeV---the soft processes
may follow different rules. In case of  a semihard 
quark or gluon exchange the outgoing partons will be
either colour triplet or octet sources of  radiation
respectively, so
in the soft limit we expect the ratio as in (\ref{rgq}) for both cases.  

Let's now consider specifically 
the DIS process studied at HERA and construct $I_0(y)$ as
above.
Analogous results are obtained for $\gamma\gamma$ collisions.
 In the Breit frame we have a quark in the current direction recoiling
against another colour triplet system in proton direction. So in this frame
we have $I_0(y_{Breit})=I_0^q$, the density as in \epem annihilation. 
However, the same DIS process can also be initiated by other subprocesses,
for example, the photon gluon fusion. This process can be viewed 
best in a frame closer to the proton: a $q\bar q$ pair 
moves in current direction and
the remainder in proton direction. This process is mediated by gluon
exchange, so the colour octet sources yield a soft particle rate 
$I_0^g(y)=\frac{9}{4}I_0^q$ in such a frame. The overall  picture
then looks as in Fig. 2. 
\begin{figure}[t]
\begin{center}
\mbox{\epsfig{file=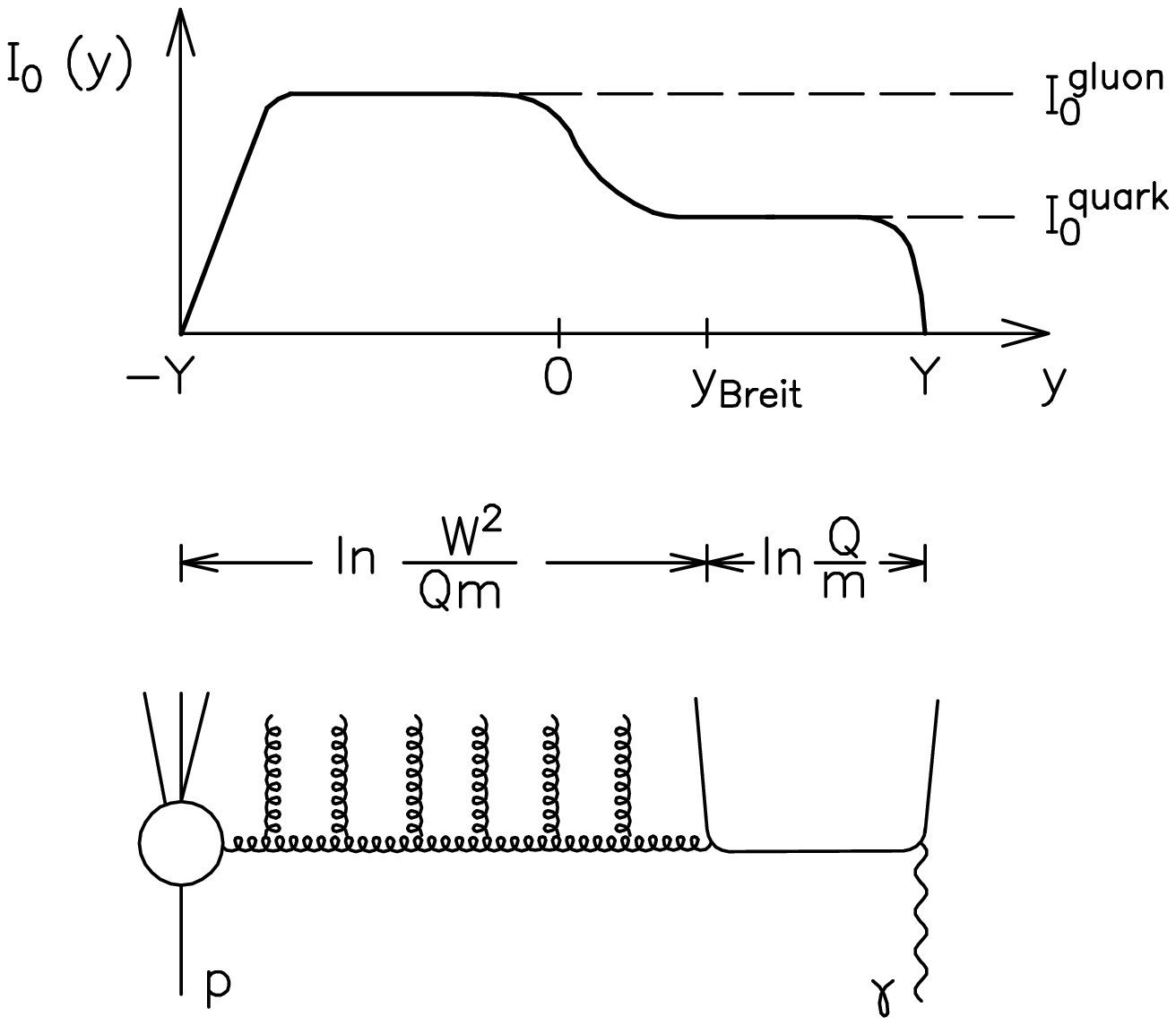,width=8.0cm,bbllx=2.8cm,bblly=12.0cm,%
bburx=16.3cm,bbury=23.7cm}}
\end{center}
\fcaption{Schematic view of particle density at small $p_\perp$ 
as a function of rapidity $y$ in the $cms$:
 in frames at rapidity $y$ with dominant gluon exchange 
the soft particle density $I_0(y)$ is 9/4 times larger  than
in the frames with dominant quark
exchange (the Breit frame, for example).} 
\label{fig:yplat}
\end{figure}

As already mentioned the exchanged parton should be sufficiently hard.
The gluon virtuality cannot be simply enforced by the external conditions
but it should become larger with increasing $Q^2$. 
We have no explicit calculation for this effect in the moment, so the
conditions are formulated at a qualitative level.
However, there are quantitative asymptotic expectations and predictions
which could be falsified. We consider 
the distribution of particles in rapidity and at
small $p_\perp$, at best in the Breit frame.
An increase of $Q^2$ should make the small $k_\perp$ 
current plateau longer but not higher than $I_0^q$. 
At the same time the plateau in proton direction 
should be rising and approach for large $Q^2$ and $W^2$ a  
 limit, the  \lq\lq gluon plateau''
  $I_0^g$, about two times larger than 
the \lq\lq quark plateau'' $I_0^q$.\\

\noindent{\it Soft radiation perpendicular to the production plane 
in multi-jet events}\\
The collinear parton configurations just discussed correspond to
quark or gluon exchanges with the soft radiation expected in the ratio
\eref{rgq}. We consider now the configurations not exactly collinear
but with moderate transverse momentum, so that the corresponding parton jets
can be resolved. In that case we consider the soft particle
production in direction perpendicular to the production plane which can be
analytically calculated in the soft region taking into account the cut-off
$Q_0$.

The simplest case is the process $\epem \to q\bar q g$. The soft radiation
in this process has been analysed already some time ago\cite{adktdrag} and applied
to the case of multiplicity flow in the production plane. The spectrum
perpendicular to the plane (say, in a small cone) is conveniently normalized
to the corresponding flow in a 2 jet $q\bar q$ event by
\begin{equation}
R^a_{\perp}(p) \equiv \frac{dN_{\perp}^{a}/d\Omega_{\vec n}dp}
     {dN_{\perp}^{q\bar q}/d\Omega_{\vec n}dp}
\label{rperpdef}
\end{equation}
for a general process $a$. For the case of $q\bar q g$ events one finds
for the soft bremsstrahlung in order $\alpha_s$\cite{klo1}
\begin{equation}
R^{q\bar q g}_{\perp}(p) =
\frac{N_C}{4 C_F} \left[ 2 - \cos \Theta_{qg} - \cos \Theta_{\bar q g} -
\frac{1}{N_C^2}
(1 - \cos \Theta_{q\bar q} ) \right].
\label{rperpqqg}
\end{equation}
where $\Theta_{ij}$ are the angles between the jets in the $cms$.
This formula yields the two 
collinear limits $R_\perp=1$ for collinear or soft gluons and
$R_\perp=\frac{9}{4}$ for antiparallel quarks recoiling against the gluon.
In this approximation the shape of the 
momentum spectrum does not depend on the inter-jet
angles but the absolute magnitude does.

The analogous analysis can be carried out for  photo-production of dijets
in $\gamma p$ or $\gamma\gamma$ collisions and also in $p\bar p$ 
collisions.\cite{klo2} 
One can distinguish between direct and resolved production.\cite{dijet,butter}
The former involves the soft
radiation from a $q\bar q g$ system, very much like in $\epem\to q\bar q g$,
 and for the
dominant small angle scattering corresponds to quark exchange, whereas the
latter at small angles corresponds to gluon exchange. One expects again the
ratio \eref{rgq} of the soft particle production 
in both processes at small scattering angles.

It will be interesting to find out whether  the particle density
varies with the angles  as predicted by \eref{rperpqqg} and occurs in the
predicted ratios in different processes
even down to low momenta of a few hundred MeV. This would provide a
strong argument in favour of the perturbative QCD interpretation
 of the soft particle production.

\section{Large Rapidity Gaps in \epem Annihilation}
 In the previous example of inclusive particle spectra
the partons and hadrons are typically close in phase space
and the hadronization requires only small rearrangements of momenta.
If during the evolution of the parton cascade in \epem annihilation
a large rapidity
gap occurs, one may expect the colour confinement forces 
at large distances to fill the gap by hadrons in the hadronization phase.
Then there is no close relation between parton and hadron final states.
The study of large rapidity gap events can give us therefore 
further insight into the colour neutralization mechanism.

The interest in such events has been awakened recently by the findings
of large rapidity gaps in dijet events at TEVATRON\cite{tevagap} and 
HERA\cite{heragap} which are interpreted by the exchange of colour neutral
objects.

In \epem annihilation---if one wants to create a large gap without colour
exchange such as to avoid final state rearrangements by confinement forces%
---one is lead to hard processes 
of the type $e^+ e^- \to q\bar q q\bar q$
or $q\bar q gg$ where the hadrons are produced from colour singlet low mass 
$q\bar q$ or $ gg$ clusters.\cite{bbh,er} As these processes 
in the perturbative analysis involve a highly virtual
intermediate quark or gluon
their rate is
rather small though. One finds that the rate keeps decreasing
with increasing gap size, contrary to the case of $p \bar p$ or $ep$
collisions.
 
Indeed, the recent measurement of rapidity gaps by 
the SLD collaboration\cite{sld} 
has shown an unlimited decrease of the gap rate 
over five orders of
magnitude (such a decrease has also been seen at lower energies 
some time ago\cite{hrs}). However, the gap rate in absolute terms
exceeds the expectations from the above calculations\cite{bbh,er}
by about two orders of magnitude and therefore, this type of hard 
colour neutralization cannot  be the dominant process for the formation
of large rapidity gap events.

An alternative approach,\cite{os} 
in line with the previously discussed LPHD,
considers a rather soft confinement mechanism, i.e. the hadron formation
does not change the parton final state considerably, not even if a large
rapidity gap is formed. Clearly, such an
application of LPHD cannot be justified a priori.  
Rather one tries and looks how different the rates of large gaps 
really are in the parton and hadron cascades.

One has to calculate the probability that no parton is emitted into a
certain angular interval, say  
between $\Theta_1$ and $
\Theta_2$ ($\Theta_1 < \Theta_2$) where $\Theta$ is
measured with respect to the initial parton direction. 
The rapidity for a massless parton is then obtained
from $y = -\ln {\rm tg} \frac{\Theta}{2}$. 
This probability is given by the
so-called Sudakov formfactor. For the actual problem of the 
angular ordered cascade with transverse momentum cut-off it has been
calculated in connection with the multi-jet rates in the
Durham/$p_\perp$ algorithm.\cite{cdotw,do}

The probability that no parton be emitted below the angle $\Theta$
in the cascade with cut-off $p_\perp>Q_0$ initiated by the primary parton
$A$ of momentum \Pj
is given by
\begin{eqnarray}
\Delta_A(\Pj, \Theta,Q_0) &=& \exp (-w_A(\Pj,\Theta,Q_0)) \label{Delta} \\
w_A(\Pj,\Theta,Q_0) &=& \int d\omega^\prime \int_{k_\perp>Q_0} d\Theta^\prime
\wp_A(\omega^\prime,\Theta^\prime)
\label{exactdoubleintegration},
\end{eqnarray}  
where $ \wp_A(\omega^\prime,\Theta^\prime)=
dn_A/d\omega^\prime d\Theta^\prime$ is the emission density of a gluon at
energy $\omega^\prime$ and angle $\Theta^\prime$. 
The probability for a gap between the two angles 
$\Theta_1 < \Theta_2 < \frac{\pi}{2}$ is then given by the ratio
of two Sudakov formfactors
\begin{eqnarray}
f_A(\Theta_1,\Theta_2)=\Delta_A(\Theta_2)/\Delta_A(\Theta_1)
= \exp \left( -w_A(\Theta_2)+w_A(\Theta_1) \right).
\end{eqnarray}
 
In the simplest approximation, the DLA, one finds
\begin{equation}
w_A(Y,\lambda)=\frac{C_A}{N_C}\beta^2
\{   
(Y+\lambda)\ln \frac{Y+\lambda}{\lambda}-Y
\}   
  \label{wdla}
\end{equation}
In this approximation the jets evolve independently in the two hemispheres
and the respective gap probabilities factorize. In particular, 
for the symmetric gap in \epem annihilation one obtains accordingly
\begin{equation}
f_A(\Theta_G)=e^{-2(w_A(\frac{\pi}{2})-w_A(\Theta_G))}.   \label{fsym}
\end{equation}
A result with improved accuracy is obtained by taking into account the exact
form of the matrix element for $\epem\to q\bar q g$
with numerical evaluation of the integral 
\eref{exactdoubleintegration}.\cite{os} An analytical approximation in MLLA
can be obtained by simplifying the integral over the nonsingular parts
of the splitting functions.\cite{do,os}

As the result of these calculations one finds the probability for the 
gap in the symmetric rapidity interval $\Delta y\approx
-2\ln\frac{\Theta_G}{2}$. It shows an almost exponential decrease very much
like the experimental data. The slope is not very different for DLA and MLLA
calculations but strongly depends on the 
parameter $\lambda=\ln(Q_0/\Lambda)$.                                                   

\begin{figure}[t]
\begin{center}
\mbox{\epsfig{file=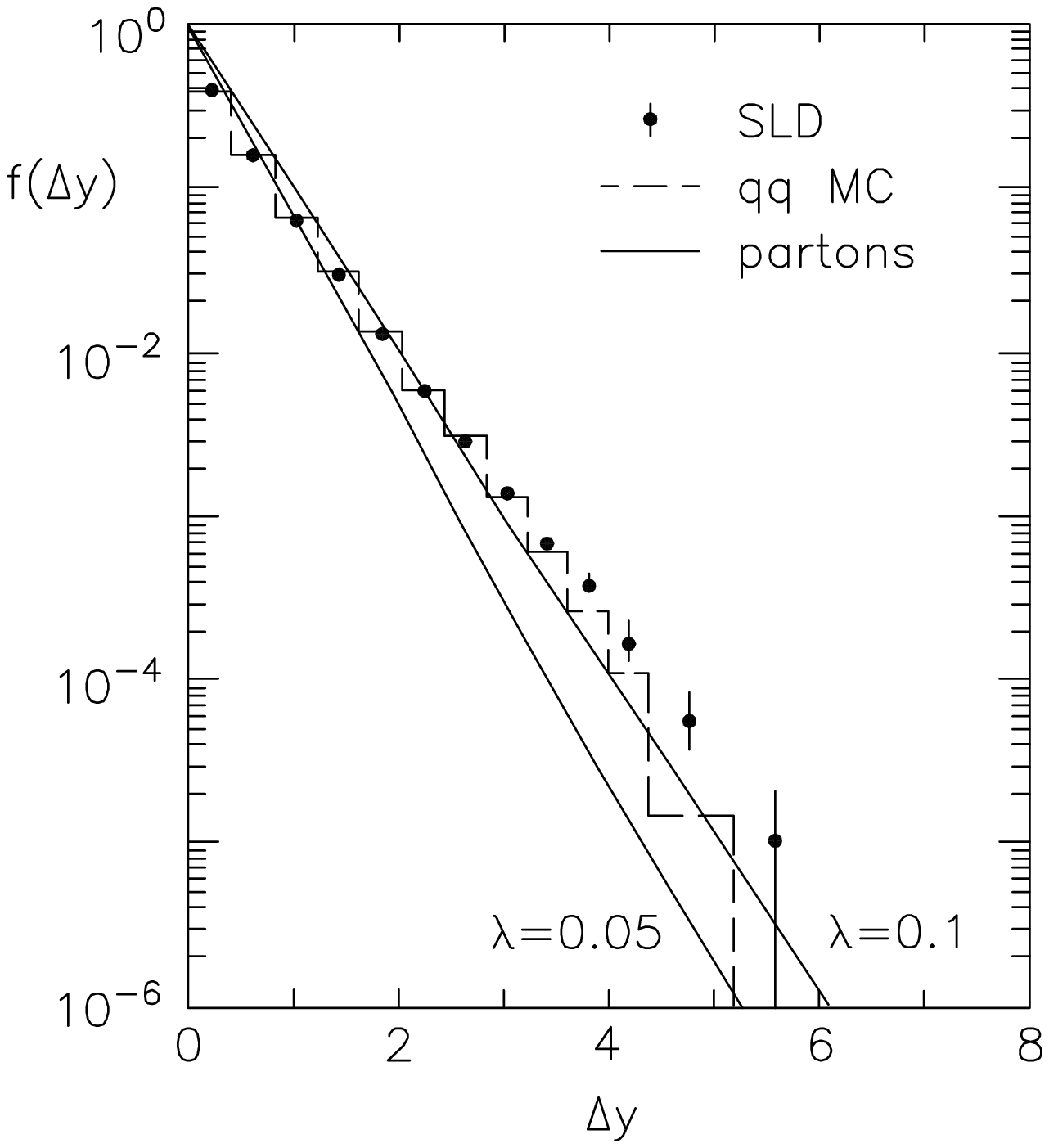,bbllx=3.0cm,bblly=9.0cm,%
bburx=16.5cm,bbury=23.7cm,height=8cm}}
\end{center}
\fcaption{Fraction of rapidity gap events in $e^+e^-$ annihilation   
as a function of the full width of 
the symmetric gap. The data points refer to the measurement of
gaps between charged particles ($\tau$-lepton events included) by
the SLD Collaboration.\protect\cite{sld} The dashed histogram shows the
expectation from the JETSET \protect\cite{JETSET} Monte Carlo without
$\tau$-leptons. Also shown are the DLA predictions for the gaps of a parton
cascade for two values of $\lambda=\ln(Q_0/\Lambda$ at $Q_0=0.244$ GeV.}
\end{figure} 

The SLD data\cite{sld} are shown in Fig. 3. They refer to rapidity gaps
between charged particles and still include $\tau$-lepton events.
The analytical calculations\cite{os} from the DLA are 
shown for two parameters of
$\lambda$ at $\Lambda=0.244$. 
An upper limit $\lambda<0.1$ has been derived 
from a moment analysis of the spectra.\cite{lo} 
These calculations should rather be compared to
data on gaps empty of any  particles, not only of charged ones.
Their distribution is expected to have a steeper slope than
shown by the data in the figure. 
As an example the prediction for $\lambda=0.05$ is shown. 
A recent analysis of
particle and jet multiplicities beyond MLLA\cite{lo2} has given the small
value $\lambda= 0.015\pm 0.005$. 

One can conclude that the distribution of gaps in a hadron cascade is well
represented by the 
 one in the parton cascade, provided the cut-off $Q_0$ is taken small
enough, i.e. very close to the QCD scale $\Lambda$ in the present
approximation scheme. If one took the larger cut-off $Q_0\sim 1$ GeV
the gap probability at $\Delta y=6$ would be larger by 4 orders of
magnitude! This suggests that the hadronization phase in a conventional
hadronization model with shorter parton cascade has a similar effect as the 
evolution of the parton cascade  from $Q_0=1$ GeV down to $Q_0=0.25$ GeV.
This is not too surprising as it is in this last phase where the running
$\alpha_s$ increases most strongly and makes the suppression of radiation
increasingly difficult.  

An interesting prediction from this approach concerns the distribution of
rapidity gaps in single quark and gluon jets. As the exponent of the Sudakov
formfactor is derived from the one-particle-density it is proportional again
to the colour factor $C_A$, i.e. the slope of the gap distribution
is correspondingly steeper (roughly twice) for gluon jets than for quark jets.
This prediction could be checked in high $p_\perp$ jets at the TEVATRON or
at HERA.

 \section{Conclusions}
The description of detailed properties of the hadronic jets  in hard
processes in terms of a parton cascade assuming a duality
between parton and hadron production 
continues to be successful. It is surprising
that this duality works even for observables where a priori there is no good
reason to believe in a perturbative approach.



\begin{figure}[t] 
          \begin{center}
\mbox{\epsfig{file=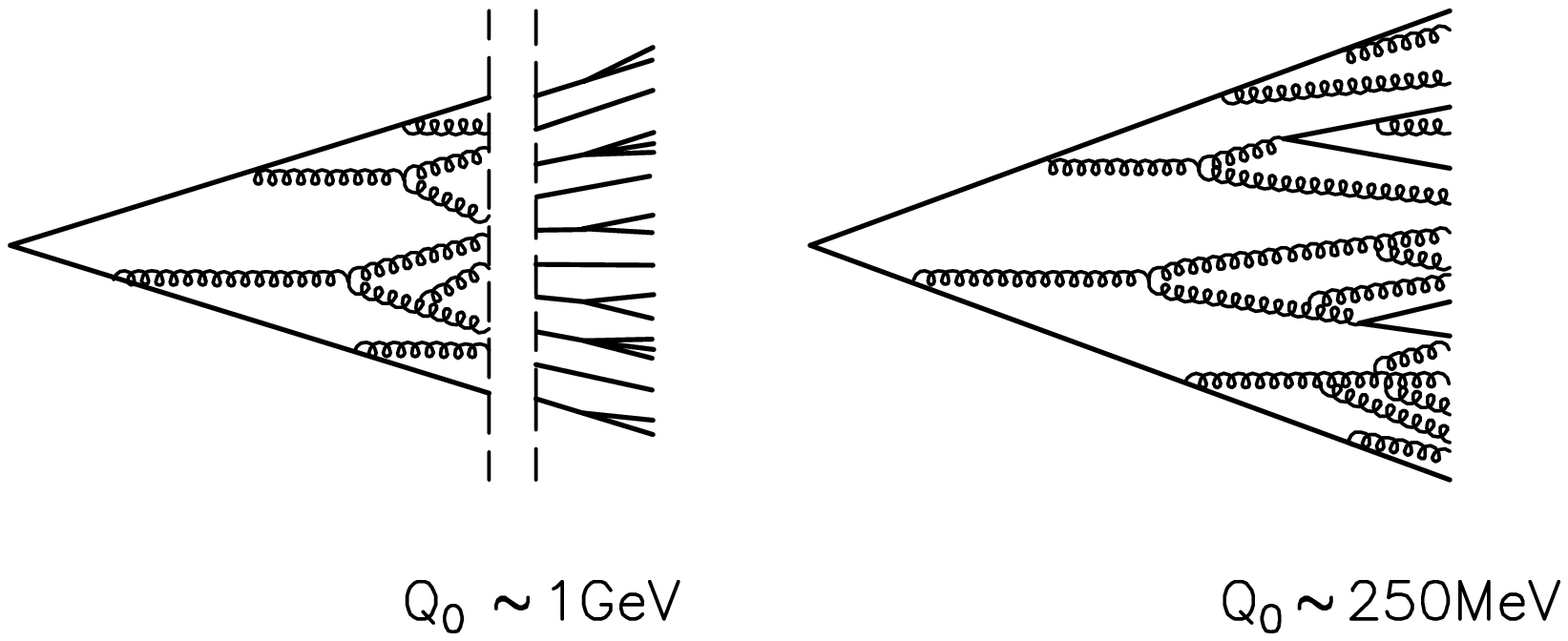,width=12.0cm,bbllx=1.8cm,bblly=14.3cm,%
bburx=19.0cm,bbury=21.8cm}}
       \end{center}
\fcaption{In the standard hadronization model (left) the parton cascade is
followed by a hadronization phase. In the dual approach the hadronization
phase is
assumed to be represented by the parton cascade in an average sense after
evolution towards a 
small cut-off scale $Q_0$}
\label{fig:radk}
\end{figure}

It appears 
that the evolution of the parton cascade at very low scales below 1 GeV
resembles in its main inclusive aspects the hadronization phase of the
phenomenological models with all its resonance decays,
 so one can speak of a dual description
of the hadronic phenomena after appropriate averaging (see Fig. 4).
As the cut-off $Q_0$ is close to the QCD scale $\Lambda$
within less than 10\% the coupling $\alpha_s$
becomes large in the last phase of the perturbative cascade.
So it may not be unreasonable that the strong coupling perturbative regime
corresponds to the hadronic phase with resonance production.

The QCD expectation of the energy independence of the low momentum particle
production from the soft gluon coherence is well met by the hadronic data in
the full energy region explored, i.e. in \epem annihilation 
from ADONE at 1.6 GeV up to LEP-1.5 and in DIS from $Q^2=12$ GeV$^2$ to
$Q^2=100$ GeV$^2$.
It will be interesting to test the predictions 
on the higher production rates of soft particles
for other colour configurations than $q\bar
q$ which can be obtained in various processes.

The parton cascade with low cut-off $Q_0$ and without hadronization
also reproduces the approximately exponential decrease of 
the rapidity gap distribution in \epem annihilation. This suggests in
particular that the confinement of colour in the jet evolution is a soft
process.

The correspondence between the parton and hadron final state 
according to these results is not like the conventional duality
between a colour singlet $q\bar q$  and a hadronic cluster.
Rather it is a correspondence between a coloured parton (typically a gluon)
and a colourless hadron. These partons are taken at a particular scale
$p_\perp>Q_0\gsim \Lambda$, so that partons below that scale are not resolved.
  These soft non-perturbative  sea-partons could take the important role
in the confinement process without disturbing the momenta of the 
resolved perturbative partons. 

Meanwhile it seems to be a challenge to test further this dual picture
of hadron formation at the limit of what is accessible perturbatively.

\section{Acknowledgements}
It is a pleasure to thank 
V.A. Khoze, S. Lupia and T. Shimada for the collaboration on the topics
discussed in this talk. 
Thanks also to B. Kniehl and R. Saffert for having the 
meeting so well organized.


\section{References}
\begin{thebibliography}{99}
\bibitem{JETSET}   
T.\ Sj\"{o}strand and M.\ Bengtsson, {\it Comp.\ Phys.\ Commun.}
{\bf 43} (1987) 367.

\bibitem{HERWIG}   
G.\ Marchesini, B.\ R.\ Webber, G.\ Abbiendi, I. G. Knowles, M. H. Seymour,
and L. Stanco,
{\it Comp. Phys. Comm.} {\bf 67} (1992) 465.

\bibitem{adkt}   
 Ya.\ I.\ Azimov, Yu.\ L.\ Dokshitzer, V.\ A.\
Khoze and S.\ I.\ Troyan, {\it Z.\ Phys.} {\bf C27} (1985) 65 and
{\bf C31} (1986) 213.

\bibitem{dfk1}  
 Yu.\ L.\ Dokshitzer, V.\ S.\ Fadin and V.\ A.\
Khoze, {\it Phys.\ Lett.} {\bf 115B} (1982) 242; 
{\it Z.\ Phys.} {\bf C15} (1982) 325.

\bibitem{bcm}   
 A.\ Bassetto, M.\ Ciafaloni and G.\ Marchesini,
{\it Phys.\ Rep.} {\bf C100} (1983) 201.

\bibitem{MLLA}
 A.\ H.\ Mueller, {\it Nucl.\ Phys.} {\bf B213}
(1983) 85; 
Erratum quoted ibid., {\bf B241} (1984) 141;\\
 Yu.\ L.\ Dokshitzer and S.\ I.\ Troyan, {\it Proc.\
19th Winter School of the LNPI}, Vol.\ 1, p.144; Leningrad
preprint LNPI-922 (1984).

\bibitem{dkmt}   
 Yu.\ L.\ Dokshitzer, V.\ A.\ Khoze, A.\ H.\
Mueller and S.\ I.\ Troyan, {\it Basics of Perturbative
QCD}, ed.\ J.\ Tran Thanh Van (Editions Fronti\'{e}res,
Gif-sur-Yvette, 1991).

\bibitem{ko} 
for a recent discussion of the phenomenological status, 
see V. A. Khoze and W.~Ochs,
{\it Int.~J.~Mod.~Phys.}~{\bf A12} (1997) 2949.

\bibitem{lo2}
S. Lupia and W. Ochs, \lq\lq {\it Unified QCD Description of Hadron and Jet
Multiplicities''}, 
MPI-PhT/97-46, July 1997 (hep-ph/9707393).

\bibitem{klo1}
V. A. Khoze, S. Lupia and W. Ochs, {\it Phys. Lett.} {\bf B394} (1997) 179;\\
{\it Proc. 7th Int. Workshop on Multiparticle Production}, Eds. R.C. Hwa,
W. Kittel, W.J. Metzger and D.J. Schotanus, June 1966, Nijmegen, The
Netherlands (World Scientific, Singapore, 1997), p. 358 (hep-ph/9610348).

\bibitem{os}
W. Ochs and T. Shimada, {\it \lq\lq Rapidity Gaps in \epem Annihilation and
Parton-Hadron Duality''}, {\it Proc. 
33rd Eloisatron workshop}, Erice, Italy, Oct.
1996, to be publ. (hep-ph/9706485)

\bibitem{AO}   
 B.\ I.\ Ermolayev and V.\ S.\ Fadin, {\it JETP.\
Lett.} {\bf 33} (1981) 285;\\
 A.\ H. Mueller, {\it Phys.\ Lett.} {\bf 104B} (1981)
161.

\bibitem{do}
Yu.L.~Dokshitzer and M.~Olsson, {\it Nucl.~Phys.}~{\bf B396} (1993) 137.

\bibitem{woz}
W. Ochs, {\it Acta Phys. Pol.} {\bf B27} (1996) 3505.

\bibitem{DKTInt} 
 Yu.\ L.\ Dokshitzer, V.\ A.\ Khoze and S.\ I.\
Troyan, {\it Int. J. Mod. Phys.} {\bf A7} (1992) 1875.

\bibitem{lo} 
S. Lupia and W. Ochs, {\it Phys. Lett.} {\bf B365} (1996) 339;\\
MPI-PhT/97-26 (hep-ph/9704319) April 1997, {\it Z. Phys.} {\bf C} in press.

\bibitem{bcmm}   
 A.\ Bassetto, M.\ Ciafaloni, G.\ Marchesini and
A.\ H.\ Mueller, {\it Nucl.\ Phys.} {\bf B207} (1982) 189. 

\bibitem{H1}   
 H1 Collaboration:  C. Adloff et al., {\it \lq\lq 
Evolution of $ep$ Fragmentation
and Multiplicity Distributions in the Breit Frame''}, DESY 97-108
(hep-ex/9707005).

\bibitem{klo2}
V.A. Khoze, S. Lupia and W. Ochs, paper in preparation.

\bibitem{bg}   
 S.\ J.\ Brodsky and J.\ F.\ Gunion, {\it Phys.\ Rev.\
Lett.} {\bf 37} (1976) 402. 

\bibitem{adktdrag}
 Ya.\ I.\ Azimov, Yu.\ L.\ Dokshitzer, V.\ A.\
Khoze and S.\ I.\ Troyan, {\it Phys.\ Lett.} {\bf B165} (1985) 147.    

\bibitem{dijet}
ZEUS Collaboration: M. Derrick et al., {\it Phys. Lett.} {\bf B348} (1995) 665; 
   {\bf B384} (1996) 401.\\
H1 Collaboration: S. Aid et al., {\it  Z. Phys.} {\bf C70} (1996) 17.

\bibitem{butter} 
J.M. Butterworth, these proceedings.

\bibitem{tevagap}
D0 collaboration: S.\ Abachi et al., {\it Phys.~Rev.~Lett.}~{\bf 72} (1994)
2332.\\
CDF collaboration: F.\ Abe et al., {\it Phys.\ Rev.\ Lett.}~{74} (1995) 855.

\bibitem{heragap}
ZEUS collaboration: M.\ Derrick et al., {\it Phys.\ Lett.} {\bf B332} (1994)
228.\\
H1 collaboration: T.\ Ahmed et al., {\it Nucl.\ Phys.} {\bf B429} (1994) 477.

\bibitem{bbh}   
J.\ D.\ Bjorken, S.\ J.\ Brodsky and
     H.\ J.\ Lu. {\it  Phys.\ Lett.} {\bf B286} (1992) 153.

\bibitem{er}
J.~Ellis and D.A.~Ross. {\it Z.~Phys.} {\bf C70} (1996) 115.

\bibitem{sld}   
SLD collaboration: {\it Phys.~Rev.~Lett.}~{\bf 76} (1996) 115.

\bibitem{hrs}   
HRS collaboration: M.~Derrick et al., {\it Z.~Phys.} {\bf C35} (1987) 323.

\bibitem{cdotw}
S.~Catani, Yu.L.~Dokshitzer, M.~Olsson, G.~Turnock and B.R.~Webber,
{\it Phys.~Lett.}~{\bf B269} (1991) 432.


\end{thebibliography}
\end{document}